\documentclass[aps,pre,twocolumn,groupedaddress]{revtex4-1}
\usepackage{graphicx}% Include figure files
\usepackage{dcolumn}% Align table columns on decimal point
\usepackage{bm}% bold math

\usepackage{color}
\begin{document}

% Use the \preprint command to place your local institutional report
% number in the upper righthand corner of the title page in preprint mode.
% Multiple \preprint commands are allowed.
% Use the 'preprintnumbers' class option to override journal defaults
% to display numbers if necessary
%\preprint{}

%Title of paper
\title{Droplets climbing a rotating helical fiber}

% repeat the \author .. \affiliation  etc. as needed
% \email, \thanks, \homepage, \altaffiliation all apply to the current
% author. Explanatory text should go in the []'s, actual e-mail
% address or url should go in the {}'s for \email and \homepage.
% Please use the appropriate macro foreach each type of information

% \affiliation command applies to all authors since the last
% \affiliation command. The \affiliation command should follow the
% other information
% \affiliation can be followed by \email, \homepage, \thanks as well.
\author{B. Darbois Texier}
\author{S. Dorbolo}%
 \email{S.Dorbolo@ulg.ac.be}
\affiliation{\small{GRASP}, Physics Dept., University of Li\`{e}ge, B4000 Li\`{e}ge, Belgium.}%Lines break automatically or can be forced with \\

%Collaboration name if desired (requires use of superscriptaddress
%option in \documentclass). \noaffiliation is required (may also be
%used with the \author command).
%\collaboration can be followed by \email, \homepage, \thanks as well.
%\collaboration{}
%\noaffiliation

\date{\today}

\begin{abstract}
A liquid droplet is placed on a rotating helical fiber. We find that the droplet may slide down, attach or climb up the fiber. We inspect experimentally the domain of existence of these three behaviors as a function of the geometrical characteristics of the fiber, its angle relatively to the horizontal, the wetting properties of the fluid and the rotating speed of the helix. A theoretical model is proposed in order to capture the boundaries of the experimental phase diagram.
\end{abstract}

% insert suggested PACS numbers in braces on next line
% insert suggested keywords - APS authors don't need to do this
%\keywords{}

%\maketitle must follow title, authors, abstract, \pacs, and \keywords
\maketitle

% body of paper here - Use proper section commands
% References should be done using the \cite, \ref, and \label commands
\section{Introduction}

The study of the interaction between a liquid drop and a solid fiber is motivated by the fact this situation occurs in various applications such as mist collection \cite{duprat2012wetting} and fiber coating \cite{quere1999fluid}. The static shape of a liquid drop on a horizontal fiber was first considered by Carroll \cite{carroll1986equilibrium} ad then by McHale and collaborators \cite{mchale2002global,wu2006droplet}. Huang and coworkers \cite{huang2009equilibrium} inspected the problem of the equilibrium of a liquid drop on a inclined fiber due to the hysteresis of the contact line. Lorenceau \textit{et al.}  \cite{lorenceau2004capturing} studied the capture dynamics of a drop impacting a horizontal fiber. The similar problem for the case of an inclined fiber was addressed by Piroird \textit{et al.} \cite{piroird2009drops}. Gilet and collaborators \cite{gilet2010droplets,gilet2009digital} studied the dynamics of a drop sliding along a fiber paving the way to digital microfluidics on a wire. Other studies extend the limit case of a straight fiber toward more complex geometries such as conical \cite{lorenceau2004drops} or functional fibers \cite{de2012buoyant,bai2012functional}. Among all the possible shapes of fiber, helical ones are encountered in curly hairs, some plant tendrils \cite{nelson1993corkscrew} or springs. In this article, we inspect the stability and the dynamics of a liquid drop on a helical fiber put into rotation. First, the experimental behaviors of a drop on a rotating helical fiber are described. Domains for which these behaviors occur are explored depending on the helical fiber geometry and the wetting properties of the liquid. After that, a theoretical model is proposed in order to predict the transitions between the different regimes. Finally, we discuss the experimental and theoretical limitations of our approach.

%Finally, the implications of this study towards helical fibers present in Nature are drawn.  \textcolor{red}{The wetting of a fiber by a drop has been solved by Borchard \cite{brochard1986spreading}}

\section{Preliminary observations}\label{sec:experiments}

\begin{figure}[h!]
\centering
	\begin{minipage}[c]{0.5\columnwidth}
  		\centering
  		\hspace{0.5cm}(a)\\
		\vspace{0.2cm}
		\includegraphics[width=4cm]{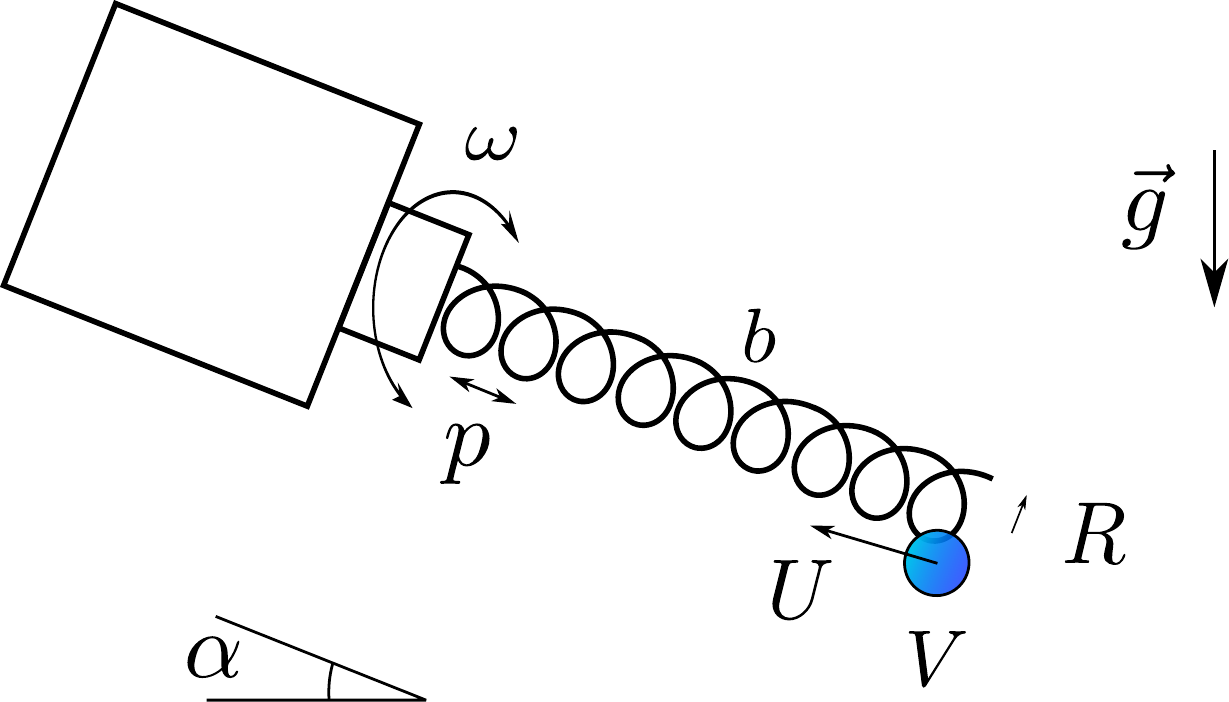}
 	\end{minipage}%
	\begin{minipage}[c]{0.5\columnwidth}
  		\centering
  		\hspace{0.5cm}(b)\\
		\vspace{0.2cm}
		\includegraphics[width=4cm]{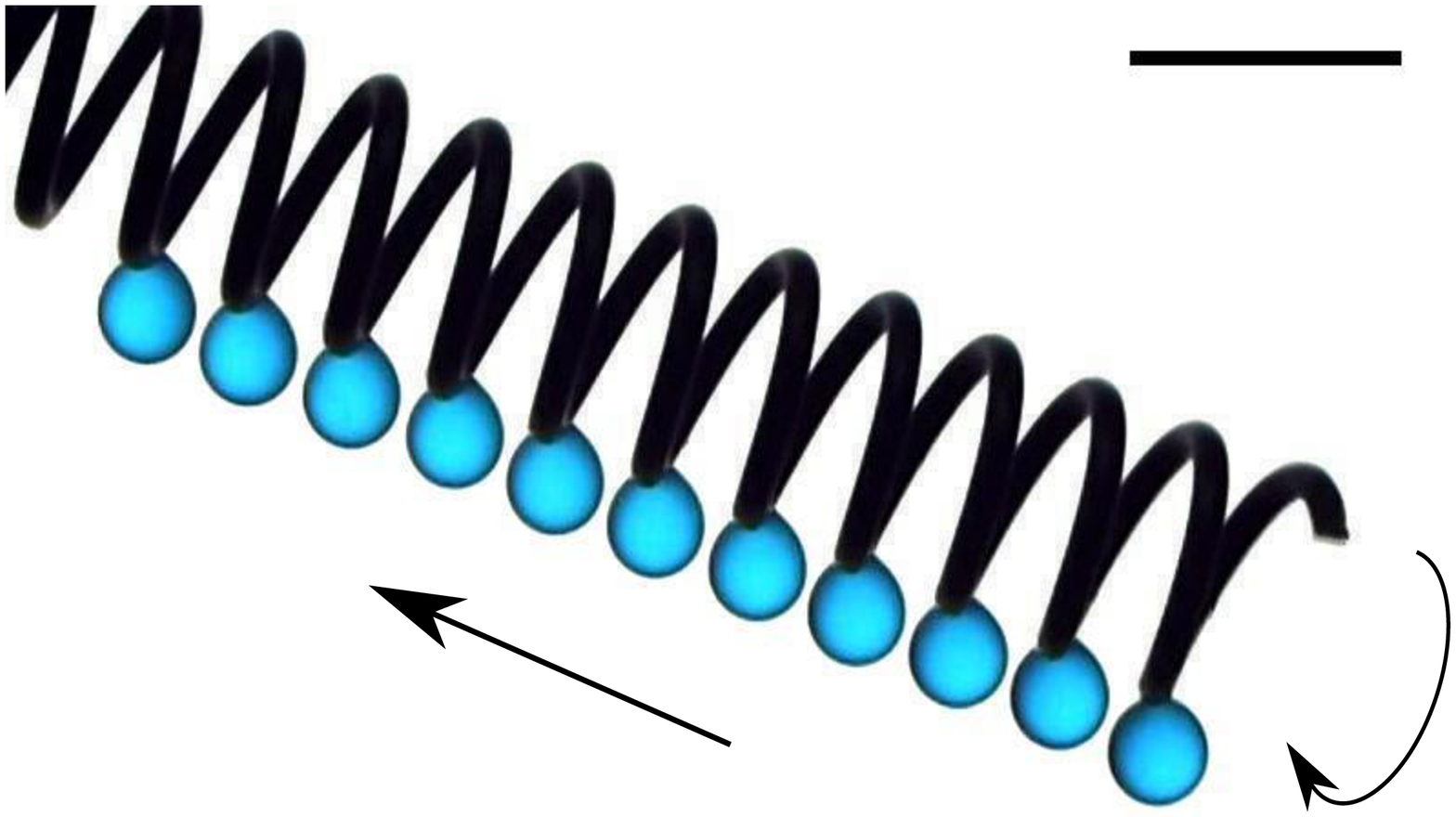}
 	\end{minipage}%
 	\caption{(a) Sketch of the experimental set-up. A liquid drop of volume $V$ is placed at the end of an helical fiber of radius $b$. The pitch of the helix is $p$ and its radius $R$. The system is put into rotation by a motor with a rotating speed $\omega$ and the speed of the drop along the helix axis is $U$. The central axis of the helical fiber is tilted by an angle $\alpha$ relatively to the horizontal and the gravitational acceleration is denoted $g$. (b) Chronophotography of a water drop pending at the end of a helix made of steel with $b=0.6 \, \rm{mm}$, $R=3.1 \, \rm{mm}$ and $p=2.9 \, \rm{mm}$ giving $\beta=8.5 \,^\circ$. The angular speed of the system is $\omega= 6.3 \, \rm{rad/s}$. The time interval between two positions is 1 s and the black line indicates 5 mm.}
		\label{fig:set-up}
\end{figure}

We attached a steel fiber with a helical shape to the rotor of a motor [Fig. \ref{fig:set-up}-(a)]. The rotating speed $\omega$ of the system can be varied from $0.1 \, \rm{rad/s}$ to $10 \, \rm{rad/s}$. The fiber has a radius $b$, the helix has a pitch $p$ and a radius $R$. The geometrical aspect of the helix defines a local slope $ \beta = \tan^{-1} (p/2 \pi R)$. This system is tilted from an angle $\alpha$ relatively to the horizontal. A liquid drop of volume $V$, density $\rho$, surface tension $\gamma$ and dynamic viscosity $\eta$ is gently placed at the end of the helical fiber and the rotating motion is started. The dynamics of the drop is recorded from the side with a camera. 

When the rotation starts, we observe situations where the drop stays at the bottom of a loop and thus follows the helix motion. If the rotation screws the helix upwards, the drop climbs the fiber [Fig. \ref{fig:set-up}-(b)]. The purpose of this article is to inspect under which conditions the drop climbing motion occurs.

% Such a behavior can be useful for applications which require control of drops motion.

\section{Completely wetting liquids}

\subsection{Equilibrium state}

\begin{figure}[h!]
\centering
	\begin{minipage}[c]{0.3\columnwidth}
  		\centering
  		\hspace{0cm}(a)\\
		\vspace{0.95cm}
		\includegraphics[width=2cm]{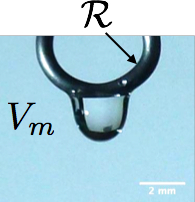}
		\vspace{0.8cm}
 	\end{minipage}%
	\begin{minipage}[c]{0.6\columnwidth}
  		\centering
  		\hspace{0.2cm}(b)\\
		\vspace{0.2cm}
		\includegraphics[width=5cm]{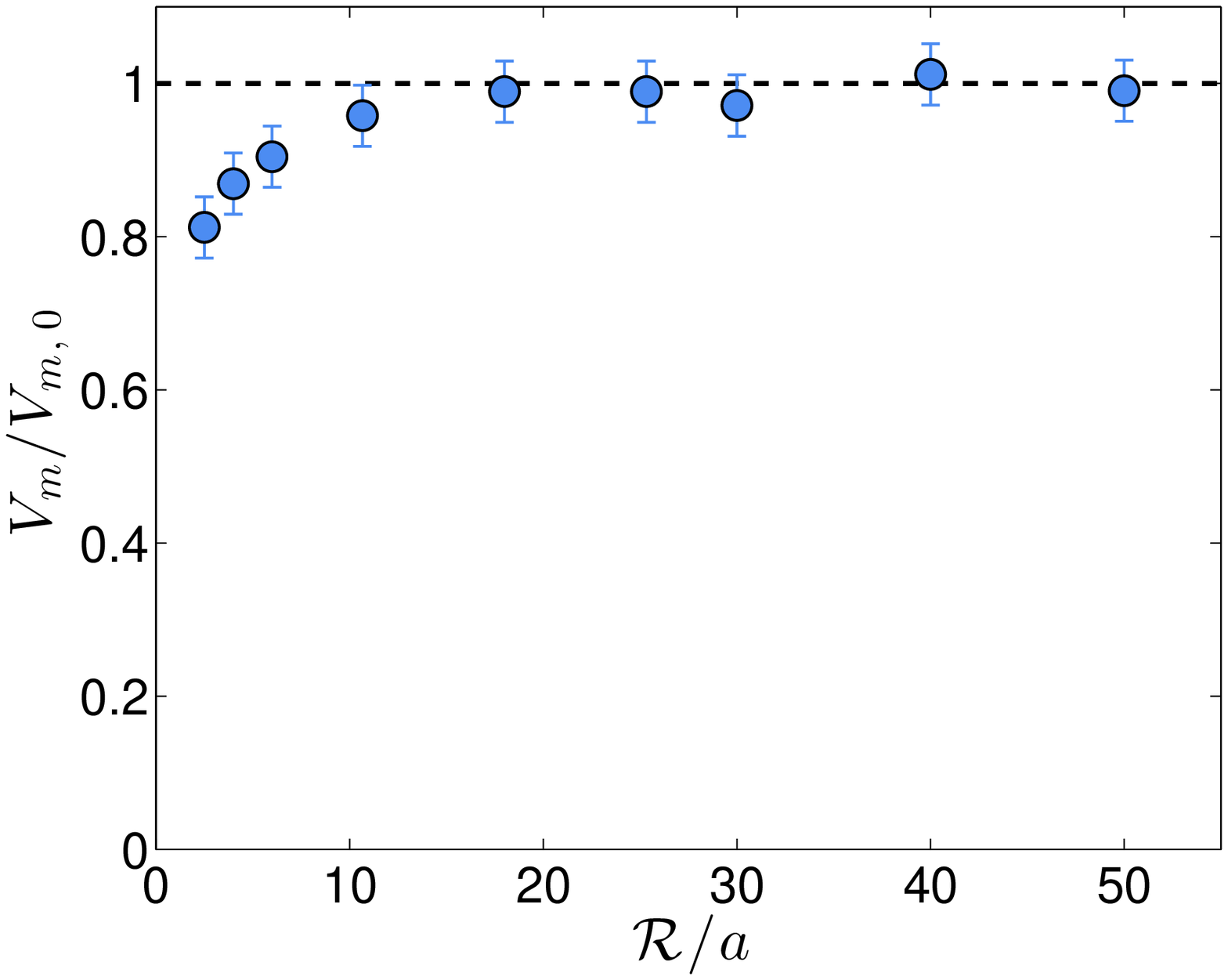}
 	\end{minipage}%
 	\caption{(a) Picture of a silicone oil drop ($V_m=8 \, \rm{mL}$, $\eta = 10 \, \rm{mPa \cdot s}$, $\rho=930 \, \rm{kg/m^3}$ and $\gamma=23 \, \rm{mN/m}$) placed on a curved fiber of radius $b=0.6 \, \rm{mm}$ and radius of curvature $\mathcal{R}=2.5 \, \rm{mm}$. (b) Evolution of the ratio $V_m/V_{m,0}$ as a function of the fiber normalized radius of curvature $\mathcal{R}/a$ for a fiber of radius $b=0.6$ mm in the case of a perfectly wetting fluid (silicone oil on a steel fiber).}
		\label{fig:fiber_curvature}
\end{figure}

The first question is to determine the maximal drop volume that can be placed on a static helical fiber without falling. In the case of a drop of a perfectly wetting fluid on a straight and horizontal fiber of radius $b$ smaller than the capillary length $a=\sqrt{\gamma/\rho g}$, Lorenceau \textit{et al.} \cite{lorenceau2004capturing} used geometrical considerations to predict that the maximal drop volume is $V_{m,0}=4 \pi \, b a^2$. We study experimentally the impact of the fiber radius of curvature $\mathcal{R}$ on the maximal equilibrium volume $V_m$ [Fig. \ref{fig:fiber_curvature}-(a)] of a silicone oil drop, a liquid which perfectly wets the steel material of the fiber. Figure \ref{fig:fiber_curvature}-(b) shows the evolution of the ratio $V_m/V_{m,0}$ as a function of the normalized radius of curvature $\mathcal{R}/a$.

% One notices that in such a situation, a clamshell configuration is preferred for the drop than a barrel shape \cite{mchale2001shape}.

One observes that for large fiber curvatures ($\mathcal{R}>15 \, a$), the maximal drop volume recovers the one $V_{m,0}$ expected in the case of a straight and horizontal fiber. For smaller fiber curvature ($\mathcal{R}<15 \,  a$), the maximal drop volume is smaller. % \textcolor{red}{Find a reference which reports a similar effect of the curvature.}

\subsection{Dynamics}\label{sec:oil_motion}

\begin{figure}[h!]
\centering
	\begin{minipage}[c]{0.45\columnwidth}
  		\centering
  		\hspace{0.1cm}(a)\\
		\vspace{0.6cm}
		\includegraphics[height=3cm]{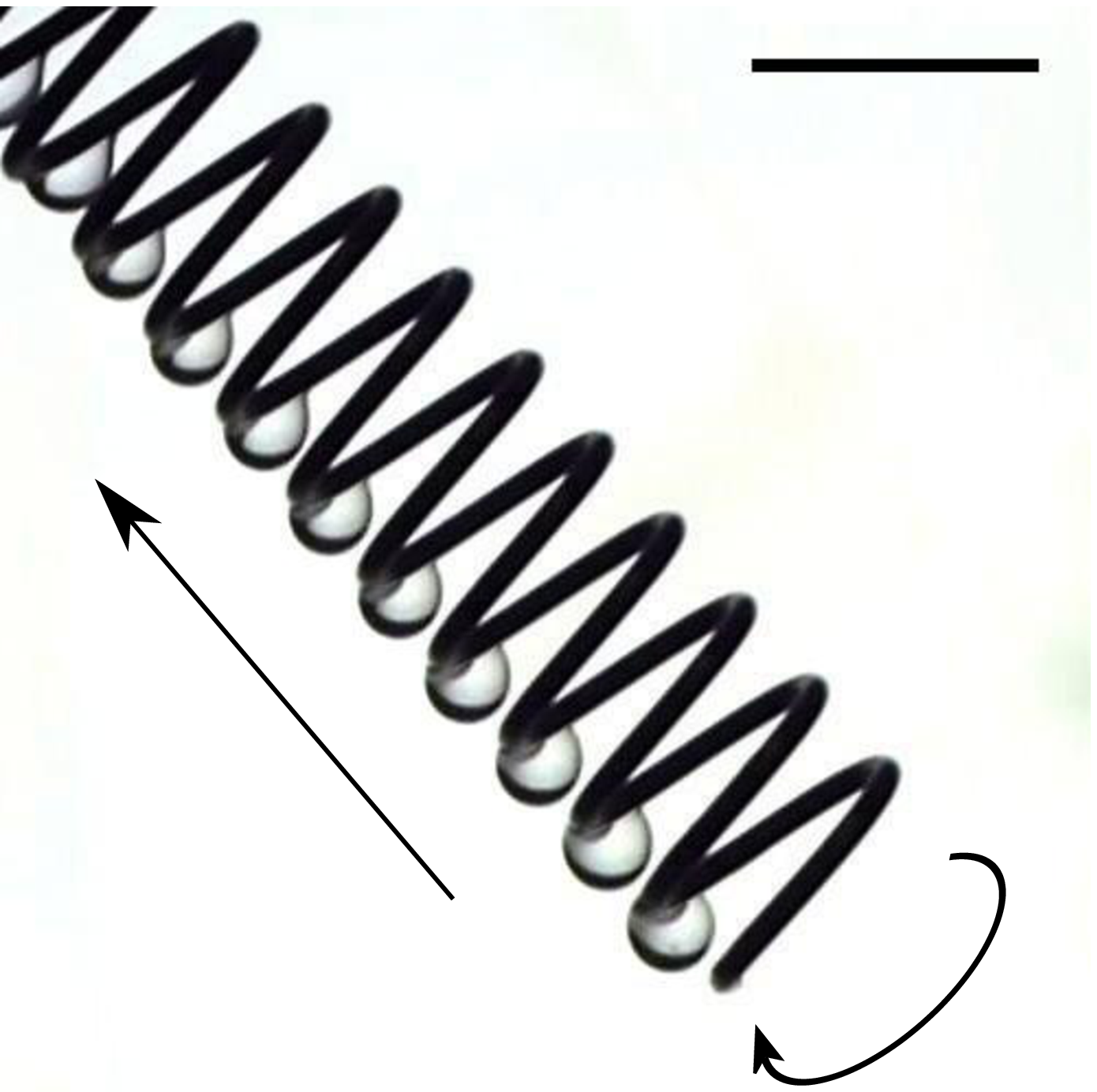}
		\vspace{0.6cm}
 	\end{minipage}%
	\begin{minipage}[c]{0.45\columnwidth}
  		\centering
  		\hspace{0.5cm}(b)\\
		\vspace{0.2cm}
		\includegraphics[height=4cm]{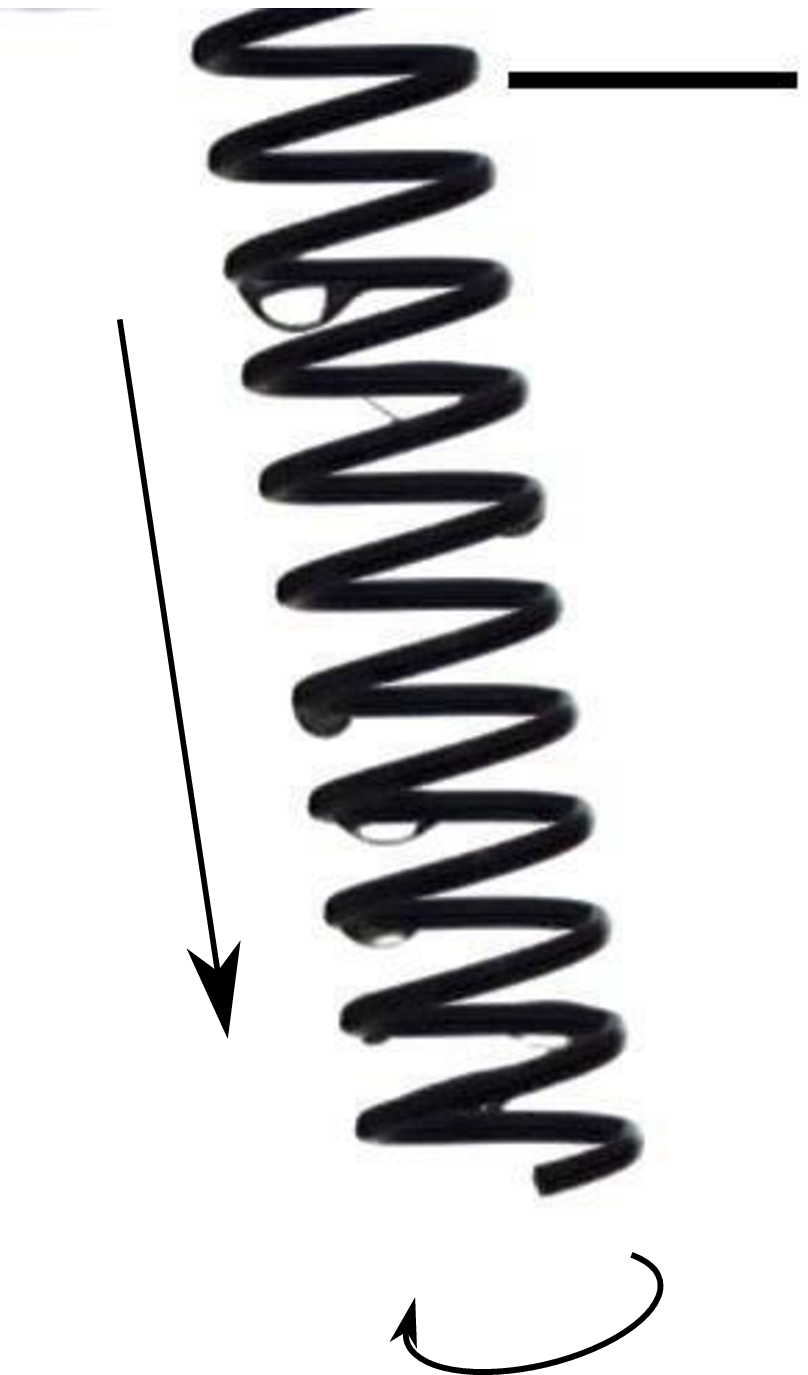}
 	\end{minipage}\\
	\vspace{0.5cm}
		\begin{minipage}[c]{0.9\columnwidth}
  		\centering
  		\hspace{0.5cm}(c)\\
		\vspace{0.1cm}
		\includegraphics[height=4cm]{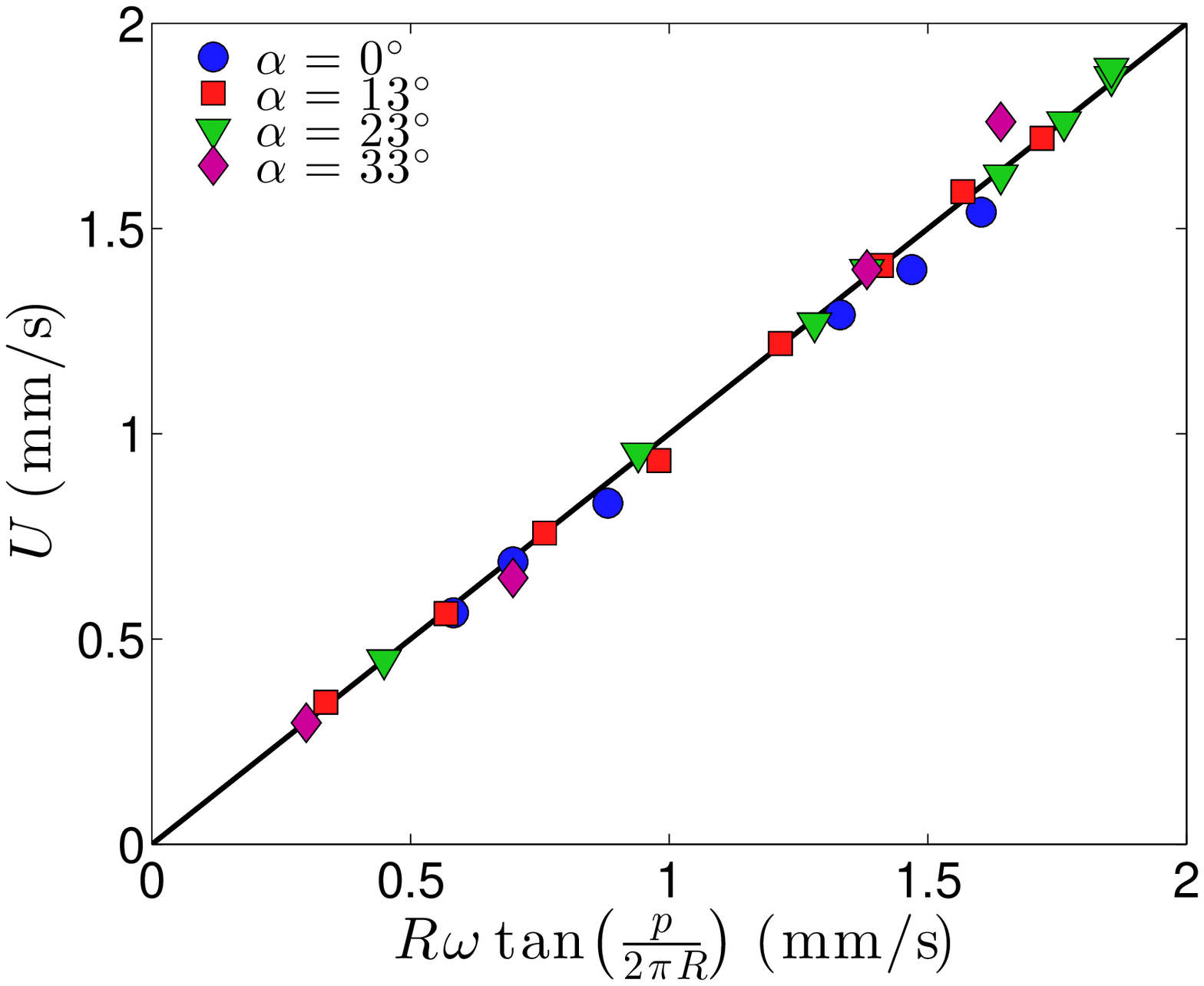}
 	\end{minipage}%
 	\caption{(a) and (b) Chronophotographies of a silicone oil drop ($\eta=10 \, \rm{mPa \cdot s}$, $\rho=930 \, \rm{kg/m^3}$ and  $\gamma=23 \, \rm{mN/m}$) placed on an helical fiber of radius $b=0.6 \, \rm{mm}$ which rotates at $\omega=3.1 \, \rm{rad/s}$. The radius of the helix is $R=3.1 \, \rm{mm}$ and its pitch $p=2.9 \, \rm{mm}$. The time interval between each picture is 2.0 s and black lines indicate 5 mm. The fiber is inclined with an angle $\alpha=49 \,^\circ$ and $\alpha=82 \,^\circ$  for respectively (a) and (b). (c) Speed $U$ of a drop along the helix axis as a function of the theoretical speed $R \omega \tan( p / 2 \pi R)$ for different angles $\alpha$ below the critical angle $\alpha_c$. Blue dots, red squares, green triangle and purple diamonds correspond respectively to $\alpha=0^\circ $, $\alpha=13^\circ $, $\alpha=23^\circ $ and $\alpha=33^\circ $.}
		\label{fig:oil_helix}
\end{figure}

This section inspects the dynamics of a silicone oil drop placed at the free end of the helical fiber which rotates. We observe two different behaviors depending on the angle $\alpha$ of the helix relatively to the horizontal. Below a critical angle $\alpha_c$, the drop stays in the minimum potential of gravitational energy formed by a loop and thus follows the helix motion. If the rotation screws the helix upwards, the drop motion is correlated to the rotating motion of the helical fiber as shown in Fig. \ref{fig:oil_helix}-(a) and the system acts as an "open" Archimedes screw at the scale of a drop \cite{rorres2000turn}. Thanks to geometrical considerations, the velocity of the bottom of a loop along the helix axis can be expressed as $R \omega \tan (p / 2 \pi R)$. Figure \ref{fig:oil_helix}-(c) shows the measured velocity $U$ of a drop along the helix axis as a function of this theoretical speed for different rotating speeds $\omega$ and different angles $\alpha$ lower than $\alpha_c$. The good agreement of the experimental and theoretical data proves that the drop perfectly slides along the fiber. 

Above the critical angle $\alpha_c$, a drop of silicone oil placed on the fiber slides downwards whatever the rotating speed $\omega$ of the helix. Figure \ref{fig:oil_helix}-(b) shows a situation where a drop, initially placed at the top of an inclined helical fiber, slides down. In such a regime, drop and helix motions are not correlated. This behavior occurs when the local slope of the helical fiber relatively to the horizontal is negative everywhere. As the drop perfectly wets the fiber, it can not sustain its own weight and it slides down. It implies that the critical angle of the helix above which a perfectly wetting drop slides down is $\alpha_c=\pi/2 - \beta$. For the helix considered in our experiments (\textit{cf.} section \ref{sec:experiments}), the theoretical value of the critical angle is $\alpha_c =81.5 \,^\circ$. This prediction is consistent with observations reported in Fig. \ref{fig:oil_helix}-(a) and (b) and has been verified experimentally within a precision of $0.2^\circ$.

\section{Partially wetting liquids}

\subsection{Dynamics}

\begin{figure}[h!]
\centering
		\includegraphics[height=6.3cm]{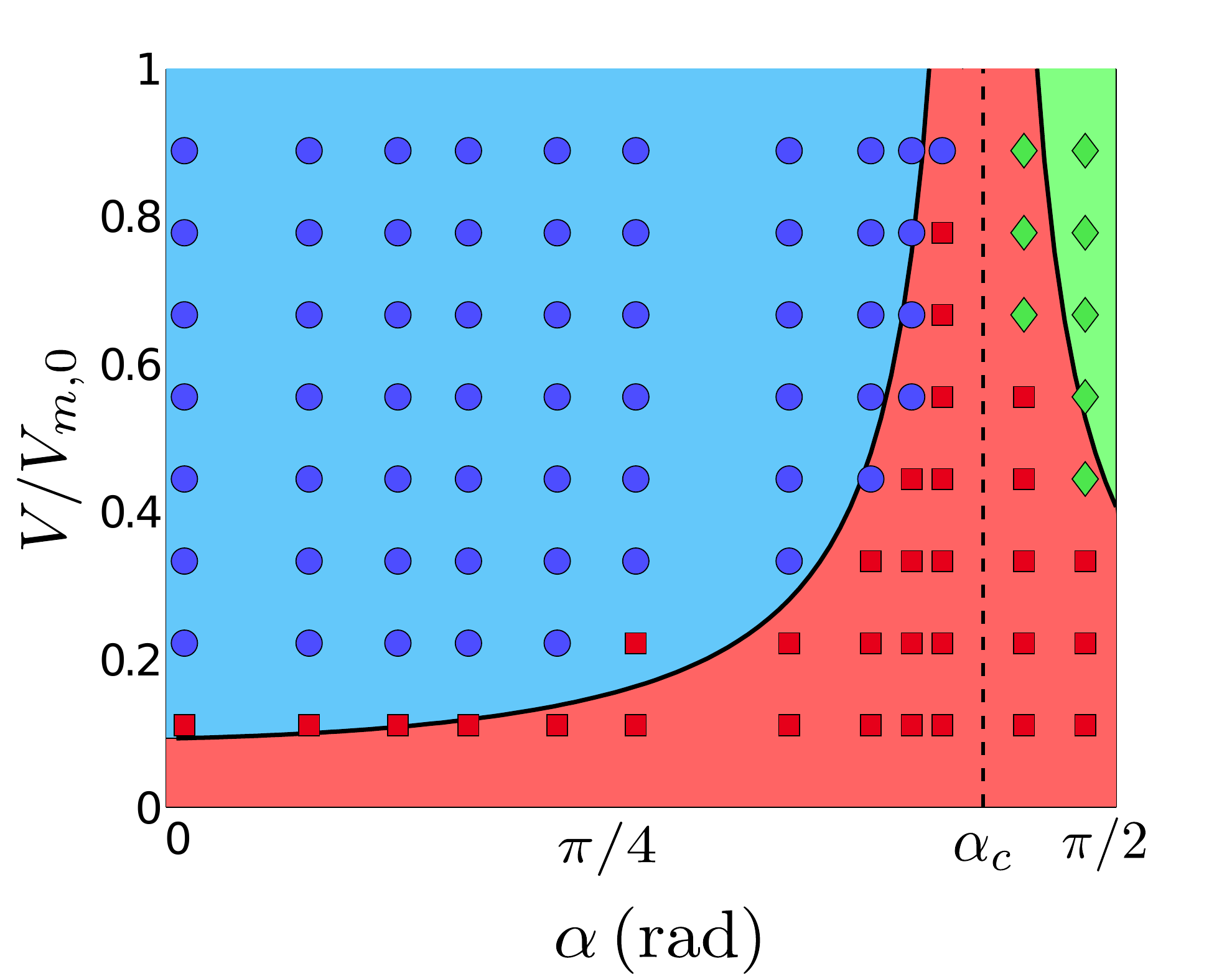}
 	\caption{Phase diagram of the behavior of a water drop  motion ($\eta = 1.0 \, \rm{mPa \cdot s}$, $\rho=1000 \, \rm{kg/m^3}$ and $\gamma= 72 \, \rm{mN/m}$) on an helical fiber ($b=0.6 \, \rm{mm}$, $R=3.1 \, \rm{mm}$ and $p=3 \, \rm{mm}$) rotating upwards ($\omega=6.3 \, \rm{rad/s}$) as a function of the helix angle $\alpha$ and the normalized drop volume $V/V_{m,0}$. Blue dots correspond to the climbing motion, green diamonds to the going down motion and the red squares to the situation where the drop is clamped at a given position. The black vertical dotted line shows the critical angle $\alpha_c$ and the black solid lines are derived from eq (\ref{eq:critical_volume_ratio_final}) for $\Delta \cos \theta = 0.57$.}
		\label{fig:phase_diagram}
\end{figure}

The same experiments as described in the section \ref{sec:oil_motion} have been conducted with deionized water ($\eta = 1.0 \, \rm{mPa \cdot s}$, $\rho=1000 \, \rm{kg/s}$ and $\gamma= 72 \, \rm{mN/m}$), a liquid which does not wet perfectly the fiber. In addition to climbing and moving downwards motions observed previously, it exists situations where the drop is attached to the fiber. In this case, the water drop does not slide along the fiber but the hysteresis of the contact line balances its own weight. We inspected experimentally the behavior of water drop placed on a rotating helical fiber depending on its angle $\alpha$ and the drop volume $V$. Results of these experiments are shown as a phase diagram in Fig. \ref{fig:phase_diagram} by the way of symbols.

One observes that the transition between the climbing and the descending motion occurs for $\alpha=\alpha_c$ as discussed previously. The water drop attaches the fiber if its volume $V$ is smaller than a critical volume $V_{m,\alpha}$ which strongly depends on the slope $\alpha$ of the helix. Closer $\alpha$ is from the critical angle $\alpha_c$, larger is the critical volume below which the drop is attached to the fiber. Besides giving rise to a third regime, the hysteresis of the contact line also induces a stick-slip motion of the drop in a bottom of a loop during a climbing motion.

\subsection{Model}\label{sec:model}

\begin{figure}[h!]
\centering
	\begin{minipage}[c]{0.45\columnwidth}
  		\centering
  		\hspace{0.1cm}(a)\\
		\vspace{0.6cm}
		\includegraphics[height=1.8cm]{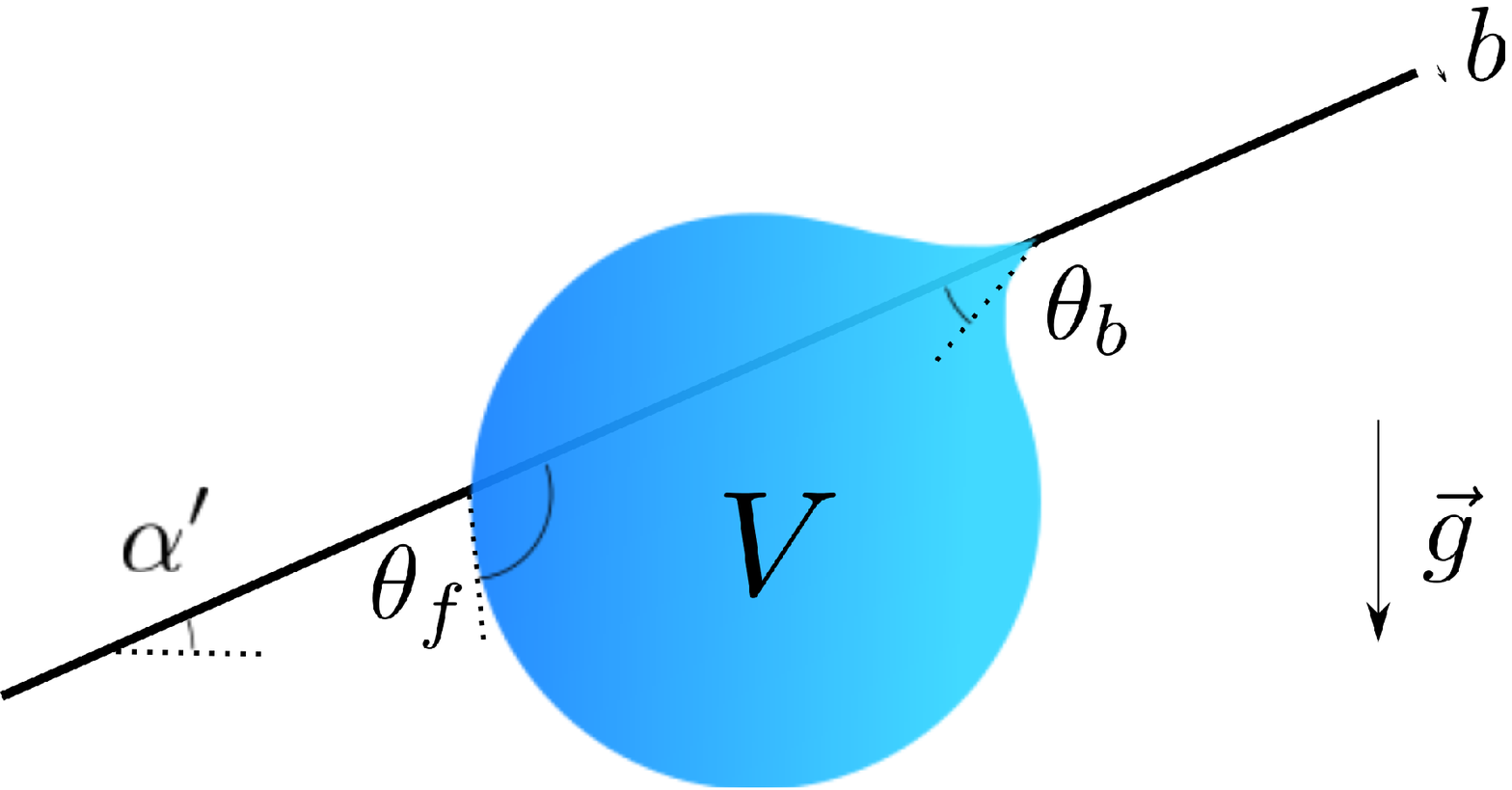}
		\vspace{0.6cm}
 	\end{minipage}%
	\begin{minipage}[c]{0.45\columnwidth}
  		\centering
  		\hspace{0.15cm}(b)\\
		\vspace{0.2cm}
		\includegraphics[height=3cm]{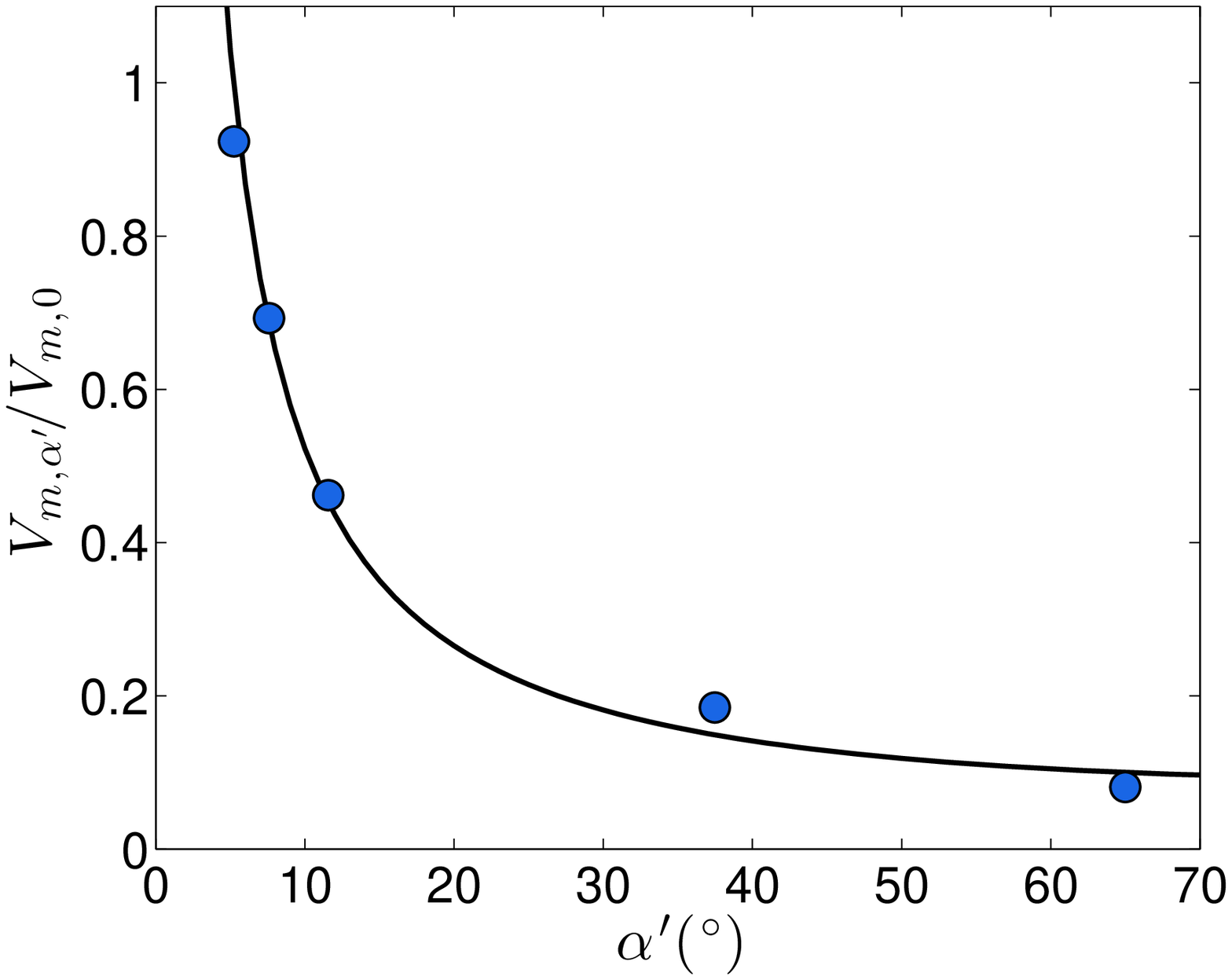}
 	\end{minipage}\\
 	\caption{(a) Sketch of a liquid drop in equilibrium on a fiber of radius $b$ tilted with an angle $\alpha'$ relatively to the horizontal. This equilibrium state is made possible by the difference between the front and back contact angles $\theta_f$ and $\theta_b$. (b) Critical volume $V_{m,\alpha '}$ below which a water drop can stand on a straight fiber tilted with an angle $\alpha'$ relatively to the horizontal as a function of the inclinaison. The fiber is made with the same material as the helical fiber in section \ref{sec:experiments} and has the same radius $b=0.6$ mm. The best fit between experimental data (blue dots) and Eq. (\ref{eq:critical_volume_ratio}) is found for $\Delta \cos \theta = 0.57$ which is represented by the the black solid line. This method allows to determine the parameter $\Delta \cos \theta$ with a precision of $\pm 0.02$.}
		\label{fig:equilibrium}
\end{figure}

%\begin{figure}[h!]
%\centering
%		\includegraphics[height=2.5cm]{figures/equilibrium}
% 	\caption{Sketch of a liquid drop in equilibrium on a fiber of radius $b$ tilted with an angle $\alpha'$ relatively to the horizontal. This equilibrium state is made possible by the difference between the front and back contact angles $\theta_f$ and $\theta_b$.}
%		\label{fig:equilibrium}
%\end{figure}

In this section, we consider the critical volume $V_{m,\alpha}$ below which a liquid drop attaches a rotating helical fiber. Huang \textit{et al.} \cite{huang2009equilibrium} studied the equilibrium of a liquid drop on a straight and inclined fiber. They showed that the hysteresis of the contact angle can induce a difference between the front and the back contact angle (respectively $\theta_f$ and $\theta_b$) of the drop on a fiber inclined of an angle $\alpha'$ [Fig. \ref{fig:equilibrium}a]. It results a force $\gamma \, 2 b \, (\cos \theta_b - \cos \theta_f)$ able to balance the horizontal component of the drop weight $\rho V g \sin \alpha '$ and lead to an equilibrium state. If $\theta+$ and $\theta_-$ are respectively the maximal front angle and the minimal back angle that the contact line can sustain, the maximal drop volume $V_{m,\alpha'}$ which is in equilibrium on a fiber tilted with an angle $\alpha'$ is given by

\begin{equation}
V_{m, \alpha'} = 2 b a^2 \frac{\cos \theta_- - \cos \theta_+}{\sin \alpha '}
\label{eq:critical_volume_straight}
\end{equation}

Introducing the parameter $\Delta \cos \theta = \cos \theta_- - \cos \theta_+$ which corresponds to the maximal contact angle hysteresis, the previous equation yields

\begin{equation}
\frac{V_{m, \alpha'}}{V_{m,0}} = \frac{1}{2 \pi} \frac{\Delta \cos \theta}{\sin \alpha '}
\label{eq:critical_volume_ratio}
\end{equation}

The parameter $\Delta \cos \theta$ is estimated experimentally by measuring the maximal angle $\alpha'$ that a drop of a volume $V_{m, \alpha'} $ can sustain before sliding along a straight and tilted fiber. Figure \ref{fig:equilibrium} shows the measurements made on a straight fiber made with the same material as the helix considered in section \ref{sec:experiments}. Fitting the data with the equation (\ref{eq:critical_volume_ratio}) allows to estimate: $\Delta \cos \theta \simeq 0.57 \pm 0.02$.

%\begin{figure}[h!]
%\centering
%		\includegraphics[height=5cm]{figures/alphap_Vm}
% 	\caption{Measurement of the critical volume $V_{m,\alpha '}$ below which a water drop can stand on a tilted fiber of inclinaison $\alpha'$}
%		\label{fig:hysteresis}
%\end{figure}

In the case of a helical fiber, the local maximal angle $\alpha'$ that a drop experiences is $\pi/2 - \alpha - \beta$. If we substitute this relation in Eq. (\ref{eq:critical_volume_ratio}), we get a theoretical estimation of the critical ratio $V_{m,\alpha}/V_{m,0}$ below which a drop does not slide on a helical fiber

\begin{equation}
\frac{V_{m, \alpha}}{V_{m,0}} = \frac{1}{2 \pi} \frac{\Delta \cos \theta}{\cos (\alpha + \beta)}
\label{eq:critical_volume_ratio_final}
\end{equation}

This theoretical boundary is plotted with dark solid lines in Fig. \ref{fig:phase_diagram}. A relative agreement between the theoretical critical volume ratio and the experimental phase domains is observed. Moreover, Eq. (\ref{eq:critical_volume_ratio_final}) predicts the critical drop volume to change between the different regimes when the helix is horizontal or vertical (respectively $\alpha=0$ and $\alpha=\pi/2$). In the case where the helix axis is horizontal, the drop follows the fiber motion if $V/V_{m,0}>\Delta \cos \theta / 2 \pi \cos \beta$. In the opposite situation where the helical fiber axis is vertical, the drop can not sustain its weight and slides down when $V/V_{m,0}>\Delta \cos \theta / 2 \pi \sin \beta$. As expected, larger the contact angle hysteresis $\Delta \cos \theta$, broader the domain where the liquid drop is attached to the helical fiber [indicated by red squares in Fig. \ref{fig:phase_diagram}]. When the geometrical angle of the helix $\beta$ increases, the critical drop volume $V_{m,\alpha}$ which follows the horizontal motion of the helix does the same whereas the critical drop volume before sliding down a vertical helix is decreased. For a given liquid and fiber material, playing with helix geometrical properties (by the way of its angle $\beta$) is a way to control drops behavior. 

\section{Discussions}

\subsection{Rotating speed}

We showed that the speed of a drop $U$ along the helix direction is proportional to $R \omega$. However, as the rotating speed of the helix is increased, the fluid deposition of the drop is enhanced. At very large rotating speed, the amount of liquid coating the helical fiber becomes non-negligible relatively to the volume of the drop. In such a situation, a drop climbing a helical fiber has a smaller volume when reaching the top of the helix and satellite droplets are visible in every loop bottom. The problem of fluid deposition on a moving fiber has been addressed by Qu\'er\'e \cite{quere1999fluid}. He showed that the thickness of the fluid deposition only depends on the capillary number which, in our context, is given by $Ca=\eta R \omega/ \gamma$. The volume of the fluid deposition during a single turn of the helix is $4 \pi^2 b R h$ with $h$ the fluid film thickness. This amount of liquid is lower than the drop volume $V$ if $h/b < V/ 4 \pi^2 b^2 R$. Considering that the drop volume is close to the maximal one $V_{m,0}$, we get $V/ 4 \pi^2 b^2 R \simeq a^2/ \pi b R \simeq 0.1$ in the case of of our experiments. The relation $h/b<0.1$ is verified if $Ca<0.01$ which means for a silicone oil drop of surface tension $\gamma = 23 \, \rm{mN/m}$ and dynamic viscosity $\eta=10 \, \rm{mPa \cdot s}$ and a helix radius $R=3 \, \rm{mm}$ that the angular speed has to verify $\omega<8 \, \rm{rad/s}$. The same consideration with the characteristics of water provides an angular speed limit of about 230 rad/s.

%\subsection{Straight and helical fibers}
%
%The comparison between a straight and a helical fiber of same diameter, composition and axis inclinaison shows that the domain where a drop slides downwards [green diamond domain in fig \ref{fig:phase_diagram}] is reduced in the second case. This observation is caused by the fact that a helical fiber which axis is tilted by an angle $\alpha$ has smaller local angles relatively to the horizontal. This difference between straight and helical fibers may have implications in nature. Indeed, some plants such as corkscrew rush (\textit{Juncus effusus spiralis}) adopt a helical shape which increases their effective bending modulus without need of additional material \cite{nelson1993corkscrew}. Plants which avoid rain drops stagnation will have an advantage towards survival because of lesser diseases development \cite{huber1992modeling}. In the case of helical plants, rain drops will slide down the ground [this regime corresponds to the green diamond domain in fig \ref{fig:phase_diagram}] if $\alpha>\alpha_c$ and $V>V_{m,\alpha}$. Larger the helix angle $\beta$, broader the domain in which the drop moves downwards. Finally, one may think that the geometrical aspect of helical structures in plant results in a compromise between their rigidity and rain drops stagnation.

\subsection{Drop shape}

In the model developed in section \ref{sec:model}, we do not consider the exact shape of the drop on the helical fiber. However, McHale \textit{et al.} \cite{mchale2001shape} and Eral \textit{et al.} \cite{eral2011drops} have shown that on a straight fiber, a drop adopts either a barrel or a clam-shell shape as a function of its volume, the contact angle and the radius of the fiber. In our situation, the transition between these two configurations may impact the length of the contact line between the drop and the fiber and thus the capillary force. Such an effect may change the prediction of the critical drop volume to slide down a tilted fiber (Eq. \ref{eq:critical_volume_straight}). Finally, the transition between a barrel and a clam-shell shape of the drop may explain the discrepancy between experiments and theory observed in Fig. \ref{fig:phase_diagram}.\\

\section*{Conclusion}

This article describes the behavior of a liquid drop on a rotating helical fiber. If the liquid perfectly wets the fiber, the drop follows the helix motion if its axis has an angle $\alpha$ relatively to the horizontal lower than a critical value $\alpha_c=\pi/2- \beta$ with $\tan \beta = p/2 \pi R$. In this regime, the system works as an Archimedes water screw and rises liquid droplets against gravity. Above the critical angle $\alpha_c$, the local slope of the helical structure is everywhere negative and the drops has no other choice than to slide downwards. If the liquid does not perfectly wet the fiber, an additional regime occurs in which the drop is attached to the fiber and stays in a given loop of the helix. This behavior is due to the hysteresis of the contact angle of the liquid on the fiber. The transition between drop attachment and the two others regimes happens for a critical drop volume $V_{m,\alpha}$ which is predicted theoretically. By tuning the helix geometry (pitch and radius) and its orientation, one is able to control the motion of a drop or to avoid its stagnation. Such a control of the motion of a liquid droplet is relevant in the context of digital microfluidic on wires \cite{gilet2009digital,weyer2015compound} and for manipulation of chemical or biological liquids which need to avoid human contact. Finally, a perspective of this work can be the study of droplets dynamics on fibers with complex geometries.
%Finally, one perspective of this study is to consider the motion of a helical fiber at the interface of a liquid pool in order to develop a system able to both extract and rise drops from this pool.\\

\begin{acknowledgments}
This research has been funded by the Inter-university Attraction Poles Programme (IAP 7/38 MicroMAST) initiated by the Belgian Science Policy Office. SD thanks the FNRS for financial support. We thank M\'ederic M\'elard and Samuel Rondia for mounting the experimental set-up. The authors acknowledge Martin Brandenbourger, Herv\'e Caps and Laurent Maquet for relevant comments on this subject.
\end{acknowledgments}

\end{document}